\documentclass{aa}  
\usepackage{natbib}
\usepackage{graphicx}
\usepackage{txfonts}
\usepackage{ulem}
\usepackage{threeparttable}

\begin{document}

   \title{Multifrequency JVLA observations of the X-shaped radio galaxy in Abell 3670}

   \author{L. Bruno
          \inst{1,2},
          M. Gitti
          \inst{1,2},
          A. Zanichelli
          \inst{2}
          \and
          L. Gregorini
          \inst{2}
          }

   \institute{Dipartimento di Fisica e Astronomia (DIFA), Universit\`a di Bologna, via Gobetti 93/2, 40129 Bologna, Italy\\
              \email{myriam.gitti@unibo.it}
         \and
             Istituto Nazionale di Astrofisica (INAF) - Istituto di Radioastronomia (IRA), via Gobetti 101, 40129 Bologna, Italy
             }


 
  \abstract
   {X-shaped radio galaxies (XRGs) exhibit a pair of bright primary lobes and a pair of weak secondary lobes ('wings'), which are oriented with an angle that gives the structure a cross-like shape. Though several theoretical models have been proposed to explain their origin, there is currently not a general consensus on a formation scenario.}
   {We analysed new multifrequency Karl G. Jansky Very Large Array (JVLA) radio data at 1.5, 5.5, 6, and 9 GHz of the candidate XRG in Abell 3670 (A3670) in order to characterise and classify it for the first time and to investigate its origin.}
   {We produced flux, spectral index, and radiative age maps of A3670 by means of the new radio data. We investigated the connection between the radio galaxy and its host, a brightest cluster galaxy (BCG) with two optical nuclei classified as a dumbbell galaxy. Finally we discussed the literature models and compared them to the observed properties of A3670.}
   {We classify A3670 as a Fanaroff-Riley I-type XRG and measured a 1.4 GHz radio power of $P_{1.4}=1.7\cdot10^{25}$ W$\cdot$Hz$^{-1}$. By estimating the radiative age of the various source components, we find that the wings are $\Delta t\simeq20$ Myr older than the lobes. We verified that the lobes and wings are aligned with the major and minor axes of the optical galaxy, respectively, and we estimated a black hole mass of $M_{\rm BH}\sim10^9 M_\odot$, which is in agreement with the typical properties of the XRGs. }
   {Among the discussed scenarios, the jet-shell interaction model may best reproduce the observed properties of A3670. The gas of a stellar shell is responsible for the deflection of the jets, thus forming the wings. The presence of stellar shells in A3670 is plausible, but it needs further optical observations to be confirmed.}

   
\titlerunning{Multifrequency JVLA observations of the XRG in A3670}
\authorrunning{Bruno et al.}
   \maketitle
%

\section{Introduction}

   Radio galaxies are morphologically classified into Fanaroff-Riley classes I (FRI) and II (FRII), depending on the lack or the presence of hotspots on the edges of their jets \citep{fanaroff}. However, there is a class of radio galaxies that does not show the typical morphology of FRI and FRII sources. These objects are called X-shaped radio galaxies (XRGs) because, apart from the bright 'primary' lobes, they exhibit a pair of weak 'secondary' lobes (or 'wings'), which are oriented with an angle that gives the structure a cross-like shape \citep{leahywilliams}. Primary lobes often host jets with hotspots, while wings never host jets.
   
    It was estimated that the XRGs represent $\sim$10\% of FRII sources \citep{leahyparma}. The sizes of the wings are comparable or longer than those of the primary lobes, even if the wings can appear smaller due to projection effects. Spectral indices are typically steeper in the wings, suggesting that they are radiatively older than the lobes. However, XRGs with a flatter index in the wings than in the lobes have been detected as well \citep{lalrao2004,lalrao2019}. Empirically, it was found that the radio power of FRIIs is higher than that of FRIs \citep{fanaroff}. The radio power of the XRGs is of the same order as FRI and FRII division power $P_{1.4} =10^{25} \; \mathrm{W\cdot Hz^{-1}}$ at 1.4 GHz \citep{dennettthorpe,cheung2009}. Moreover, radio emission seems to be linked to the host galaxy properties. Indeed, XRGs are generally associated with high ellipticity ($\varepsilon \ge 0.2$) galaxies and it was observed that the lobes and the wings are aligned with the host major and minor optical axes, respectively \citep{capetti}. Similar alignments are found in the X-rays band with the axes of the hot atmosphere surrounding the galaxy \citep{hodgeskluck2010}. A study comparing a sample of XRGs to a control sample of classical radio galaxies with similar redshift, in addition to radio and optical luminosities finds that the XRGs host supermassive black holes (SMBHs) with statistically higher masses than classical sources \citep{mezcua}.
    
   Several theoretical models have been proposed to explain the nature of XRGs and the process that generates their morphology and properties, however there is currently not a general consensus on a formation scenario \citep[see][and references therein]{gopal}. Reorientation models support the idea that wings consist of fossil emission along a previous direction of the jets \citep{wirth,merrittekers,rees,dennettthorpe,liu}. According to hydrodynamical models, the backflow plasma coming from the hotspots is diverted by a strong environment gas pressure, giving rise to the wings \citep{leahywilliams,kraft,capetti,hodgeskluck2011}. In the double active galactic nuclei (AGN) model, the lobes and wings are formed independently by two active SMBHs of a binary system \citep{lalrao2007,lalrao2019}. Finally, wings could result from the deflection of the jets after the collision with the gas present in a stellar shell \citep{gopal}.    
  
  In this work we report on our study of the radio galaxy in the Abell 3670 (\object{A3670}) galaxy cluster by means of new multifrequency Karl G. Jansky Very Large Array (JVLA) data. Previous radio observations \citep{gregorini1994} revealed the peculiar morphology of this source, making A3670 an XRG candidate. We aim to accurately study the morphology of A3670 and characterise its spectral properties for the first time in order to confirm the XRG classification and investigate the origin of its wings.
 
  The paper is organised as follows: in Sect. 2 we describe the A3670 galaxy cluster and its brightest cluster galaxy (BCG); in Sect. 3 we present the new JVLA radio data and summarise the data reduction and imaging processes; in Sect. 4 we present the radio, spectral index, and radiative age maps. We also combine the radio properties with optical properties of the BCG taken from the literature in order to discuss the theoretical models proposed to explain the XRGs origin; in Sect. 5 we compare our results to the models; in Sect. 6 we summarise our work and give our conclusions on the nature of A3670.      
  
  Throughout this paper we adopt a $\Lambda$CDM cosmology with $H_0=73\;\mathrm{km\cdot s^{-1}\cdot Mpc^{-1}}$, $\Omega_{\rm M}=0.27$ and $\Omega_{\rm \Lambda}=0.73$. The spectral index $\alpha$ is defined as $S\propto \nu^{-\alpha}$, where $S$ is the flux density, and $\nu$ is the frequency. 
  
\section{The galaxy cluster Abell 3670}  
  
  A3670 ($RA_{\rm J2000}$ $20^h14^{m}18^s$, $Dec_{\rm J2000}$  $-29^o44'51''$\footnote{http://ned.ipac.caltech.edu/}) is a richness-class 2 galaxy cluster located at redshift $z=0.142$ \citep{coziol}, which gives an angular scale of $1''=2.4 \; \mathrm{kpc}$, and with a luminosity distance of $D_{\rm L}=645$ Mpc. Its BCG is a giant elliptical galaxy classified as a dumbbell galaxy \citep{gregorini1992,andreon}. Dumbbell galaxies are optical systems presenting two nuclei with similar magnitude, which are surrounded by a common stellar halo \citep{valentijn}. The two nuclei of A3670 are separated by $7''$, corresponding to $\simeq17$ kpc according to the adopted cosmology. The BCG has an ellipticity of $\varepsilon=0.28$ and a position angle of $PA=24^o$ measured from north to east \citep{makarov}\footnote{http://leda.univ-lyon1.fr/}. Its B-band and K-band apparent magnitudes are also available: $m_{\rm B}=17.65$ and $m_{\rm K}=12.74$. We use these parameters in Sect. \ref{radio-ottico} to analyse the connection between the optical and radio properties. No X-ray data are available for this cluster.
  
  The BCG hosts a radio galaxy (\object{MRC 2011-298}) with a peculiar shape, as shown by \citet{gregorini1994}, who observed the source at 5.5 GHz using the Very Large Array (VLA). In fact, it exhibits a pair of bright lobes in the north-south direction and a pair of weak wings in the east-west direction, oriented with an angle of about 90$^o$. This morphology resembles that of the XRGs and makes A3670 a candidate of this class of objects.

\section{Observations and data reduction}
   
    \begin{table*}
   	\caption[]{Details of new JVLA radio data analysed in this work (PI: M. Gitti. Project code:14B-027).}
   	\label{dati}
   	$$
   	\begin{tabular}{ccccc}
   	\hline
   	\noalign{\smallskip}
   	&  1.5 GHz  &  5.5 GHz &  6 GHz &  9 GHz\\
   	& (L-band) & (C-band) & (C-band) & (X-band) \\
   	\noalign{\smallskip}
   	\hline
   	\noalign{\smallskip}
   	Observation date & 10-Jan.-2015 & 11-Jan.-2015 & 20-Sep.-2014 &  9-Jan.-2015   \\
   	Frequency coverage (GHz) & 1-2  & 4.5-6.5 & 4-8 & 8-10          \\
   	Array configuration & CnB & CnB & DnC & CnB \\
   	Spectral windows  & 16 & 16 & 34  & 16           \\
   	On source time (min)  & 22  & 40 & 22 & 35            \\
   	\noalign{\smallskip}
   	\hline
   	\end{tabular}
   	$$

   	\begin{tablenotes}
  	\item	{\small \textbf{Notes}. Each JVLA dataset contains a number of spectral windows, which are parted into 64 channels. Spectral windows for the datasets at 5.5, 6, and 9 GHz have 128 MHz bandwidth. Spectral windows for the dataset at 1.5 GHz have 64 MHz bandwidth.} 
   	\end{tablenotes}
   \end{table*}

    We performed new observations of the radio galaxy in A3670 with the JVLA at 1.5 (L-band), 5.5 (C-band), and 9 (X-band) GHz in CnB configuration, and at 6 (C-band) GHz in DnC configuration. The details of the observations and datasets are summarised in Table \ref{dati}. In all the observations we used 3C48 as the flux density calibrator and J2003-3251 as the phase calibrator. 
    
    Data reduction was carried out with the National Radio Astronomy Observatory (NRAO) Common Astronomy Software Applications ({\ttfamily CASA}) v. 4.7. First, we carefully edited the visibilities of the calibrators in order to remove Radio Frequency Interference (RFI), adopting both the manual and automatic flagging (with {\ttfamily RFLAG} and {\ttfamily EXTEND} modes of the {\ttfamily FLAGDATA} task). Then we performed the standard calibration procedure\footnote{The description of calibration steps can be found at https://science.nrao.edu/facilities/vla/docs/manuals/obsguide/calibration} with multiple editing iterations. Finally we self-calibrated the phases of each dataset. After the processes of calibration and self-calibration, $\sim$15\% of the visibilities in C-band (CnB configuration) and 20\% in X-band and C-band (DnC configuration) were flagged. All of them are consistent with the 15\% flagged visibilities expected at these frequencies. However, we were forced to flag more than 55\% of the L-band visibilities, due to the high level of RFI in this dataset (the typical expected value is 40\%). The flagging of entire spectral windows in L-band caused the central frequency to shift from 1.5 to 1.7 GHz.
   
  The imaging process was carried out with the {\ttfamily CLEAN} task, by setting a multifrequency synthesis mode for continuum emission analysis. The curvature of the sky was parametrised by {\ttfamily gridmode=WIDEFIELD}. In L-band, we also used {\ttfamily multiscale=[0,1,5,10]} to highlight the faint extended emission on scales from point-like to 2$\times$beam size. For each dataset, we produced three maps with different baseline weightings to study the source emission at varying resolution and sensitivity. The parameter {\ttfamily weighting=NATURAL} provides more weight to short baselines, which are sensitive to the extended emission, but it degrades the resolution. On the other hand, {\ttfamily weighting=UNIFORM} provides the same weight to short and long baselines, so sensitivity to the extended emission decreases, but the resolution improves. The parameters {\ttfamily weighting=BRIGGS, robust=N} give an intermediate weight depending on the value of the  {\ttfamily robust} parameter ({\ttfamily N=2} and {\ttfamily N=$-$2} correspond to a natural and a uniform weight, respectively). We reached the best compromise between resolution and sensitivity by fixing {\ttfamily robust=0.5}. Therefore in the following analysis we use the {\ttfamily BRIGGS=0.5} maps only (otherwise it is specified). 

\section{Results}

In this section we present the analysis of the multifrequency maps of the radio galaxy A3670. First, we show the radio, spectral index, and radiative age maps. Then, we analyse the connection between the radio galaxy and its host. 

\subsection{Radio morphology}
\label{sectionmappe}
\begin{figure*}
	\centering
	\includegraphics[height=8.5cm,width=8.5cm]{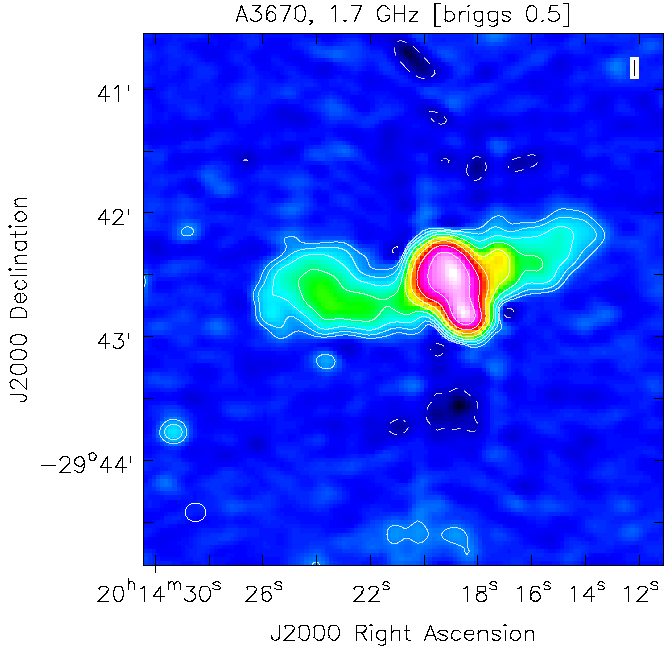} 	
	\includegraphics[height=8.5cm,width=8.8cm]{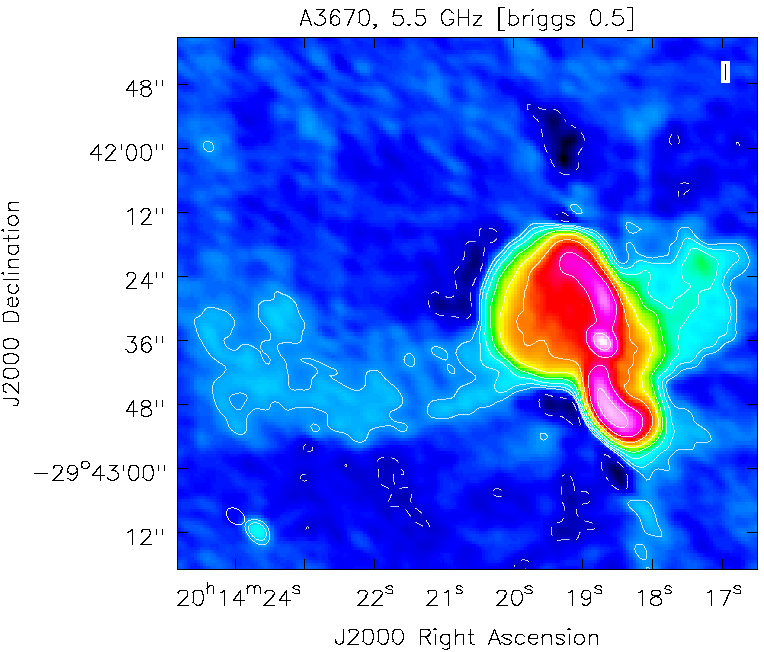}
    \includegraphics[height=8.5cm,width=8.5cm]{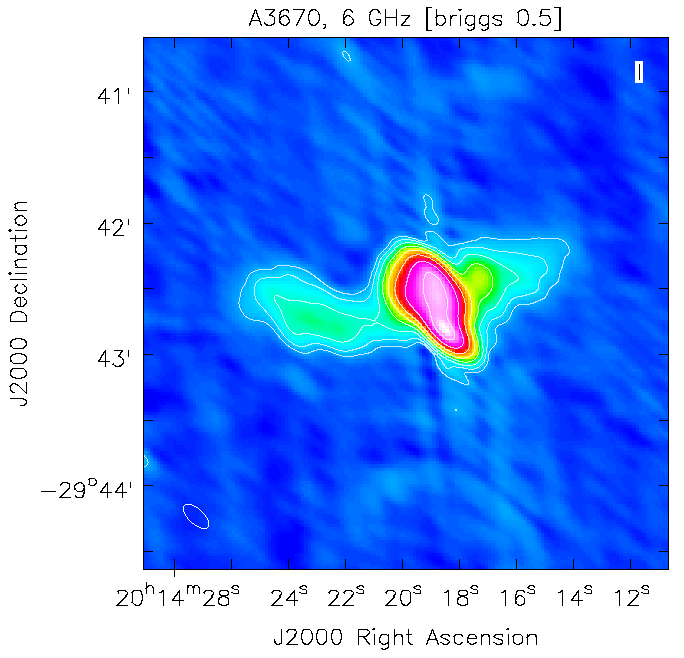} 	
	\includegraphics[height=8.5cm,width=8.5cm]{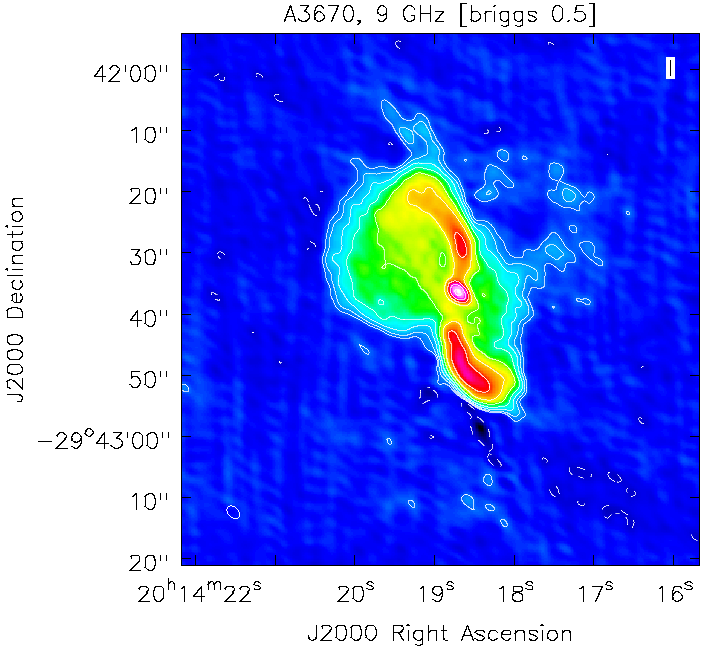} 	
	\caption{A3670 maps obtained with {\ttfamily BRIGGS 0.5} weighting parameter. In all panels, the contour levels are $-3\sigma$, $3\sigma$, $6\sigma$, $12\sigma$, $24\sigma$, $48\sigma$, ... {\it Top left}: 1.7 GHz map at a resolution $9.9''\times9.0''$ (RMS noise is 0.067 mJy$\cdot$ beam$^{-1}$). {\it Top right}: 5.5 GHz map at a resolution $3.8''\times2.7''$ (RMS noise is 0.012 mJy$\cdot$ beam$^{-1}$). {\it Bottom left}: 6 GHz map at a resolution $14.7''\times6.8''$ (RMS noise is 0.047 mJy$\cdot$ beam$^{-1}$). {\it Bottom right}: 9 GHz map at a resolution $2.5''\times1.7''$ (RMS noise is 0.007 mJy$\cdot$ beam$^{-1}$).}
	\label{mappe}
 \end{figure*}

\begin{table*}
	\caption{Properties derived from A3670 radio maps in Fig. \ref{mappe}. Columns 2 and 3 list the resolution and the RMS noise of the maps. Column 4 and 5 list the total flux density and radio power within the $3\sigma$ contour levels. Column 6 indicates the detected components and column 7 lists their flux densities.}
	\label{fluxtab}
	\begin{tabular}{ccccccc}
	\hline
	\noalign{\smallskip}
	Frequency & Resolution & RMS & Total flux density & Total radio power & Component & Component flux density  \\  
	(GHz) & (arcsec $\times$ arcsec) & (mJy$\cdot$beam$^{-1}$) & (mJy) & (W$\cdot$Hz$^{-1}$) & & (mJy)  \\  
	\hline
	\noalign{\smallskip} 
	&&&&& Primary lobes  & $294\pm15$  \\
	1.7 & $9.9\times9.0$ & 0.067 & $349\pm18$ & $(1.6\pm0.1)\cdot10^{25}$ & East wing  & $32\pm2$  \\
	&&&&& West wing & $23\pm1$ \\
	\noalign{\smallskip} \noalign{\smallskip} \noalign{\smallskip}
	&&&&& North jet  & $16.2\pm0.5$  \\
	&&&&& South jet  & $24\pm1$  \\
    5.5 & $3.8\times2.7$ & 0.012 & $127\pm4$ & $(5.9\pm0.2)\cdot10^{24}$ &Primary lobes  & $120\pm4$  \\
    &&&&& East wing  & $3.3\pm0.1$  \\
    &&&&& West wing  & $4.2\pm0.1$  \\
    \noalign{\smallskip} \noalign{\smallskip} \noalign{\smallskip}
	&&&&& Primary lobes  & $124\pm4$  \\
	6 & $14.7\times6.8$ & 0.047 & $140\pm4$ & $(6.5\pm0.2)\cdot10^{24}$ & East wing  & $8.3\pm0.2$  \\
    &&&&& West wing  & $8.4\pm0.3$  \\
	\noalign{\smallskip} \noalign{\smallskip} \noalign{\smallskip}
    &&&&& North jet  & $11.1\pm0.3 $ \\
	9 &$2.5\times1.7$& 0.007& $76\pm2 $&  $(3.5\pm0.1)\cdot10^{24}$ & South jet  & $17\pm1 $ \\
    &&&&& Primary lobes  & $76\pm2$  \\
	\noalign{\smallskip}
	\hline
	\end{tabular}
\end{table*}

A3670 maps are shown in Fig. \ref{mappe}. Table \ref{fluxtab} summarises the resolution, RMS noise, and flux density of each source component. 

The L-band map at 1.7 GHz (Fig. \ref{mappe}, top left) has a low resolution ($9.9''\times9.0''$), but it allows us to study the extended emission. The source exhibits the primary lobes along the north-south direction and a pair of weak wings, nearly perpendicular to the lobes, along the east-west direction. The flux density of the lobes is $S_{\rm lobes}=294\pm15$ mJy, while the east and the west wings have $S_{\rm Ew}=32\pm2$ mJy and $S_{\rm Ww}=23\pm1$ mJy, respectively. The total length of the lobes is $l_{\rm lobes}\simeq60''\simeq145$ kpc, whereas that of the wings is $l_{\rm Ew}\simeq75''\simeq180$ kpc and $l_{\rm Ww}\simeq60''\simeq145$ kpc. The ratio between the projected lengths of the wings and lobes is 2.8. 

Thanks to the high resolution ($3.8''\times2.7''$) reached in the C-band (array CnB) map at 5.5 GHz (Fig. \ref{mappe}, top right), we can resolve the jets and the core within the primary lobes. We observe that the jets are characterised by a curvature and an S-shaped structure. No hotspots are detected. Both the wings are faint ($S<5$ mJy), but the eastern one appears as diffuse emission, while the western wing is better defined. 
  
The resolution of the C-band map at 6 GHz (Fig. \ref{mappe}, bottom left) is lower ($14.7''\times6.8''$) and the extended emission becomes more evident. Even if we cannot resolve the jets and the core, wings are well defined. 
 
The X-band map at 9 GHz (Fig. \ref{mappe}, bottom right) has the highest resolution ($2.5''\times1.7''$) and shows well resolved jets and core, while it confirms the absence of hotspots. We notice that the jets bending acts not only in the outer regions of the source, but also in the inner ones near the core. A faint ($S\simeq1$ mJy) diffuse emission associated with the west wing is visible at the $3\sigma$ level. The south and north jets have a flux density of $S_{\rm Sj}=17\pm1$ mJy and $S_{\rm Nj}=11.1\pm0.3$ mJy, respectively. Their lengths are similar $l_{\rm Sj}\simeq l_{\rm Nj}\simeq18''\simeq40$ kpc.

On the basis of the morphological properties, we can classify A3670 as an FRI-type XRG. In order to compare the radio power of A3670 to those of classical radio galaxies and other XRGs, we measured the 1.4 GHz flux density $S_{1.4}=374\pm19$ mJy and calculated the corresponding radio power as:
\begin{equation}
P_{1.4}=4 \pi D_{\rm L}^{2}S_{1.4}(1+z)^{{\alpha-1}}=(1.7\pm0.1)\cdot10^{25} \; {\rm W\cdot Hz^{-1}} 
\label{power}
\end{equation}   
where we used the mean spectral index $\alpha=0.9\pm0.1$ between 1.7 and 9 GHz.
The radio power of A3670 is consistent with the typical XRGs radio power, intermediate between that of FRIs and FRIIs \citep{dennettthorpe}.

\subsection{Spectral indices}
 \begin{table}
	\caption[]{{\ttfamily CLEAN} parameters adopted to obtain the multifrequency maps to be used as input for the spectral index maps.}
	\label{spixtab}
	$$
	\begin{tabular}{cccc}
	\hline
	\noalign{\smallskip}
	Map 1 & Map 2  & {\ttfamily UVRANGE} & {\ttfamily RESTORINGBEAM}  \\
	(GHz) & (GHz) & (k$\lambda$) & (arcsec $\times$ arcsec) \\
	\noalign{\smallskip}
	\hline
	\noalign{\smallskip}
	5.5 & 9 & 2-91 & 3.0 $\times$ 2.0  \\ 
	1.7 & 6 & 1-23 & 12.0 $\times$ 7.0  \\ 
	1.7 & 9 & 2-23 & 9.5 $\times$ 6.5  \\   	
	\noalign{\smallskip}
	\hline
	\end{tabular}
	$$  	
\end{table}

\begin{figure*}
	\centering
	\includegraphics[height=6.8cm,width=7.8cm]{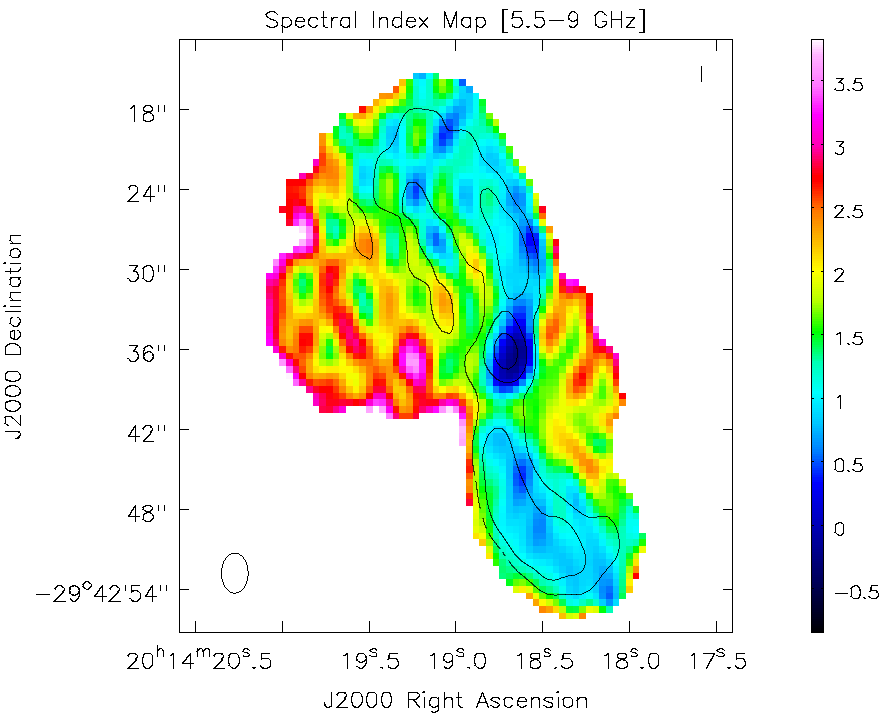}
	\includegraphics[height=6.8cm,width=7.8cm]{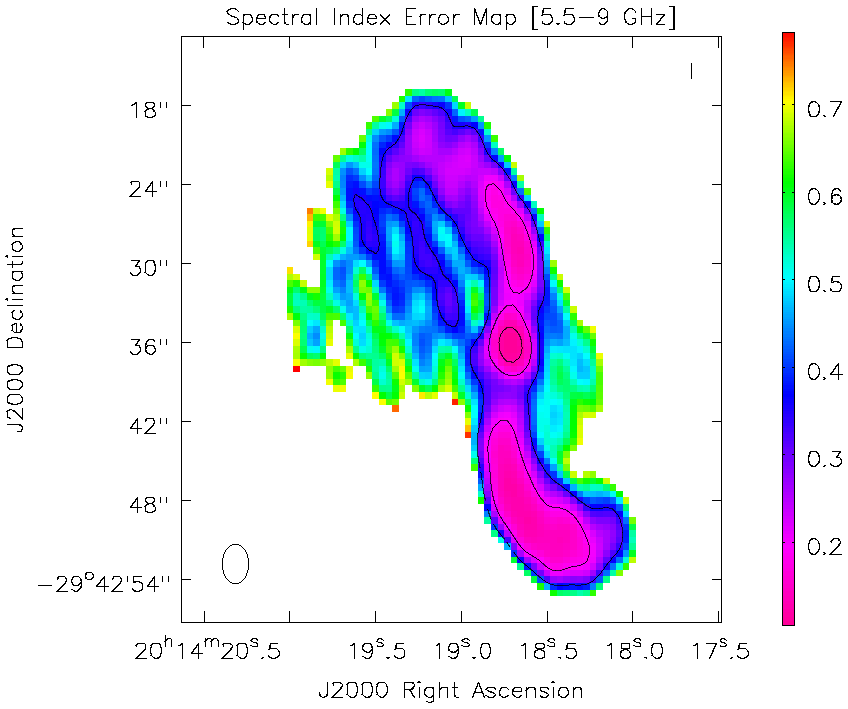}
	
	\caption{Spectral index map between 5.5 and 9 GHz at resolution $3''\times2''$ ({\it left}) and associated error map ({\it right}). Contour levels are those of the lower frequency radio map. The core has an inverted spectral index $\alpha\simeq-0.5$. Jets have $0.5\le\alpha\le1$. Primary lobes have $\alpha\simeq1.5$. The inner regions of the wings have $1.5\le\alpha\le2.5$.}
	\label{spixXC1_noL}%
\end{figure*}

\begin{figure*}
	\centering
	\includegraphics[height=6.8cm,width=7.8cm]{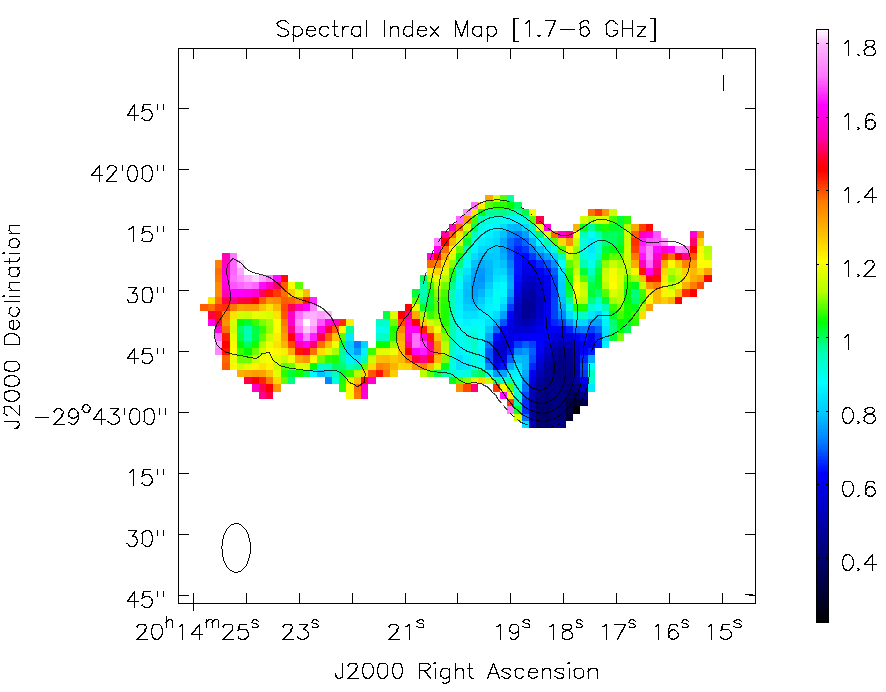} 
	\includegraphics[height=6.8cm,width=7.8cm]{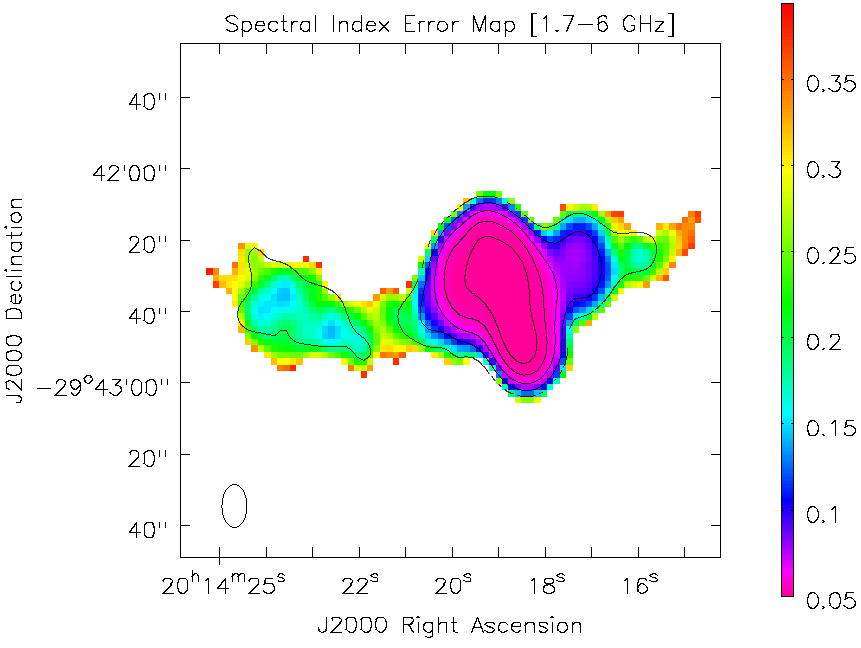}
	
	\caption{Spectral index map between 1.7 and 6 GHz at resolution $12''\times7''$ ({\it left}) and associated error map ({\it right}). Contour levels are those of the lower frequency radio map. Primary lobes have $0.5\le\alpha\le0.8$. Wings have $1\le\alpha\le1.6$.}
	\label{spixC2L}%
\end{figure*} 

\begin{figure*}
	\centering
	
	\includegraphics[height=6.8cm,width=7.8cm]{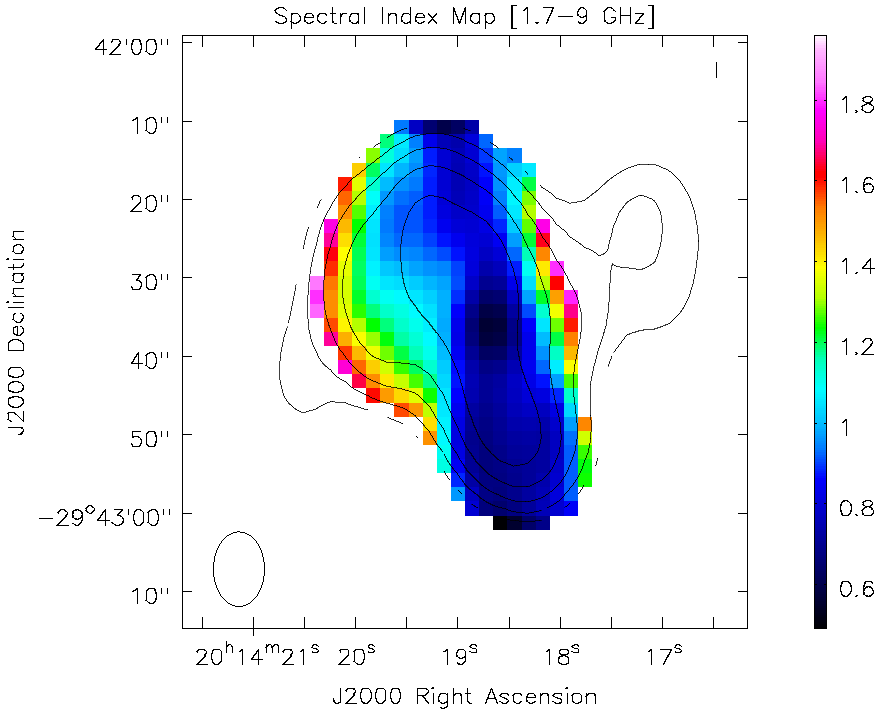} 
	\includegraphics[height=6.8cm,width=7.8cm]{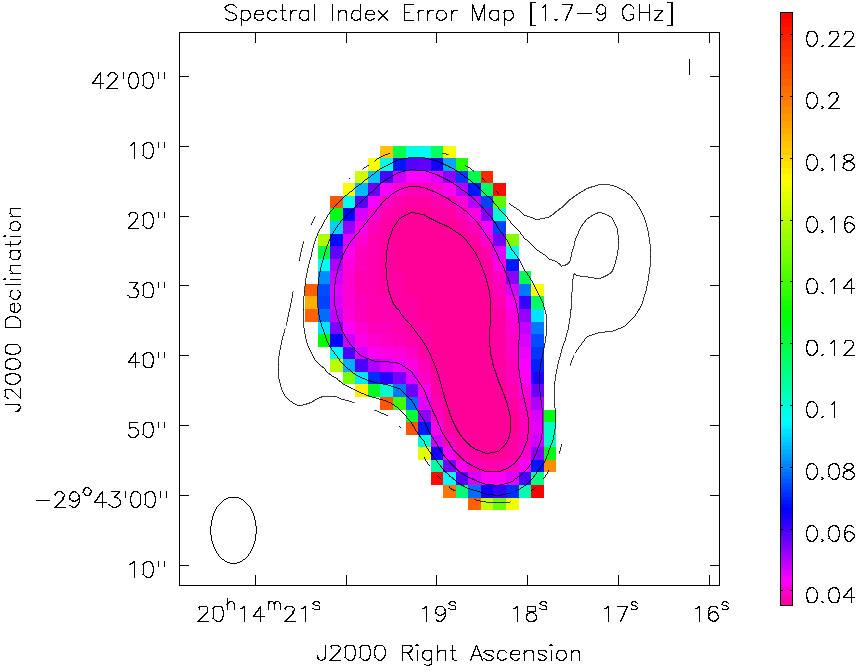}
	
	\caption{Spectral index map between 1.7 and 9 GHz at resolution $9.5''\times6.5''$ ({\it left}) and associated error map ({\it right}). Contour levels are those of the lower frequency radio map. Near the core  $\alpha\le0.6$. The jets have  $0.7\le\alpha\le0.9$. The inner parts of the wings have $1.0\le\alpha\le1.6$.}
	\label{spixXL}%
\end{figure*} 
  
Radio maps can be combined to obtain spectral index maps. For all the considered frequencies, we produced new maps with {\ttfamily weighting=UNIFORM} and the same uv range, dimensions, and clean beam. Table \ref{spixtab} summarises the adopted parameters.     
 
In Fig. \ref{spixXC1_noL} we report the spectral index map between 5.5 and 9 GHz. This high resolution map ($3''\times2''$) shows the resolved core and jets. The core exhibits an inverted spectral index of $\alpha\simeq-0.5$, probably due to self-absorption effects. The jets show a spectral index of $0.5\le\alpha\le1$. Even if the spectral index distribution is not homogenous, we notice a steepening along the east-west direction. In the primary lobes the spectral index is $\alpha\simeq1.5$, while in the inner regions of the wings it is $1.5\le\alpha\le2.5$. 

In Fig. \ref{spixC2L} we report the spectral index map between 1.7 and 6 GHz. The resolution is low ($12''\times7''$), but it allows us to study the outer parts of the wings. The jets are not resolved and in the primary lobes $0.5\le\alpha\le0.8$. The wings show a steeper spectral index than that of the lobes, with values of $1\le\alpha\le1.6$. 

In Fig. \ref{spixXL} we report the map between 1.7 and 9 GHz, which clearly shows the index steepening from the centre towards the wings. The region near the core shows a flat spectral index of $\alpha\le0.6$, in the jets $0.7\le\alpha\le0.9$ and in the outer regions $1.0\le\alpha\le1.6$.

\subsection{Radiative ages}

\begin{figure*}
	\centering
	\includegraphics[height=8cm,width=8cm]{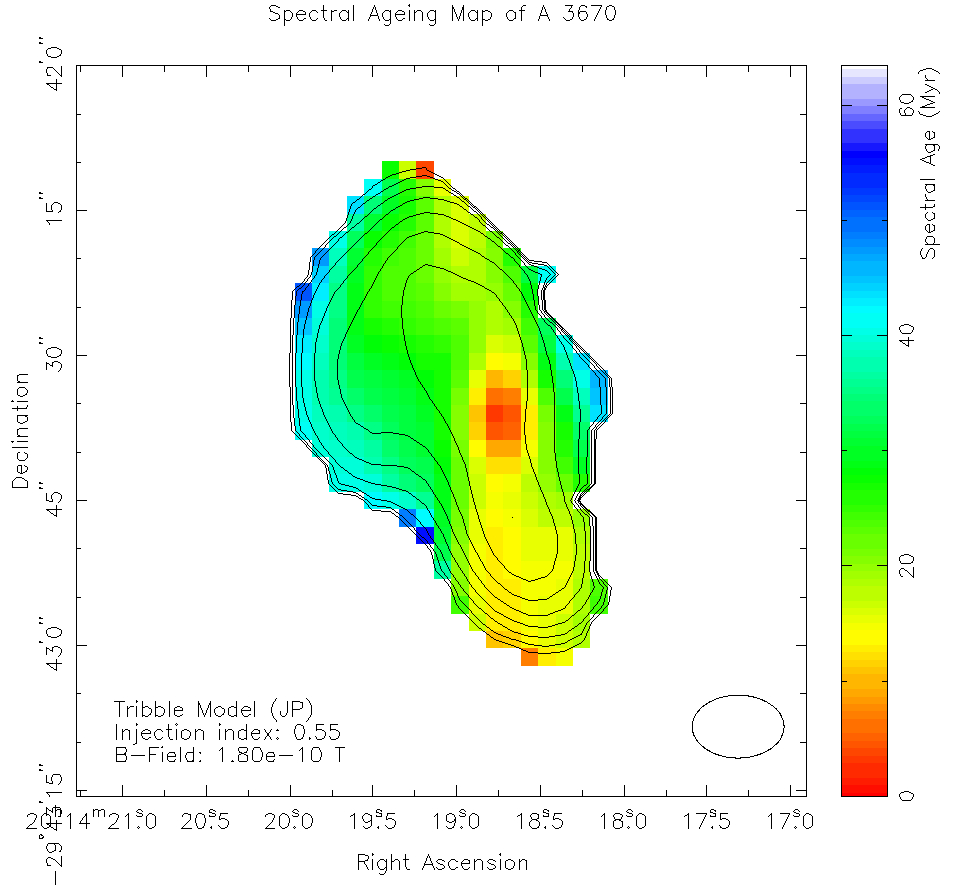} 
	\includegraphics[height=8cm,width=8cm]{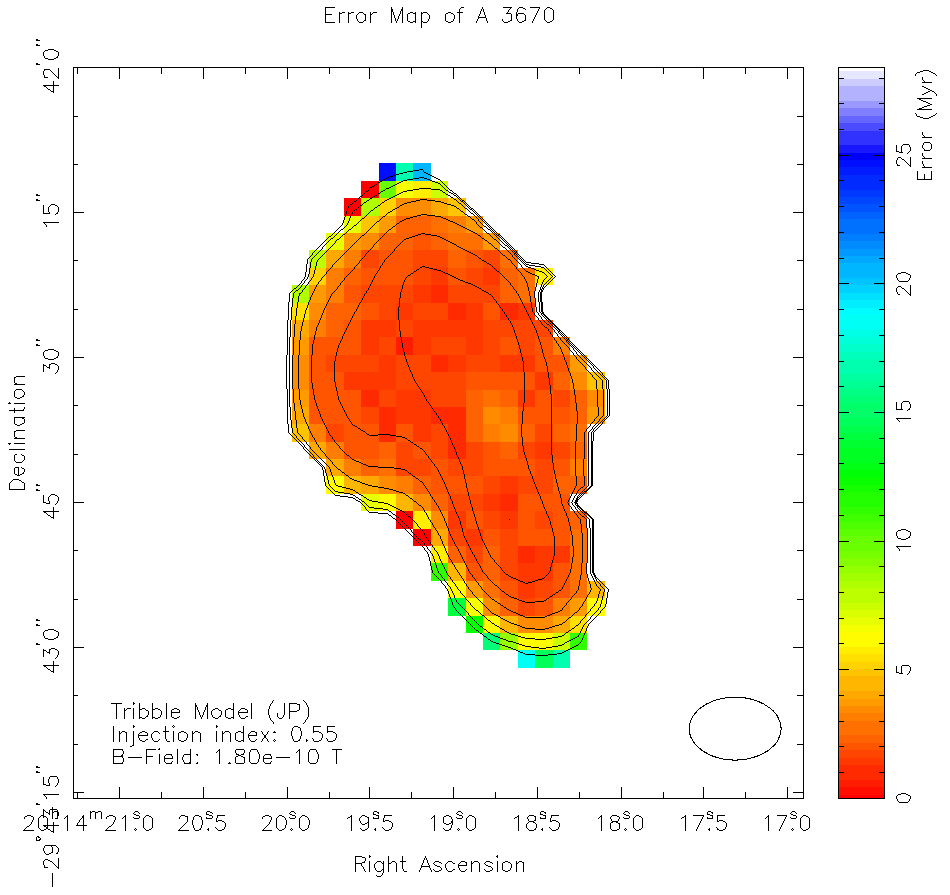}
	\caption{Radiative age ({\it left}) and associated error ({\it right}) maps obtained from the fit of a Tribble model with $B=1.8$ $\mu$G and $\Gamma=0.55$. The radiative age increases from the centre to the outer regions: the jets are $10\le t_{\rm rad}\le20$ Myr old, the primary lobes are $20\le t_{\rm rad}\le30$ Myr old, and the inner parts of the wings are $30\le t_{\rm rad}\le40$ Myr old. Typical errors are $<5$ Myr.}
	\label{agemap}%
\end{figure*} 

\begin{figure*}
	\centering
	\includegraphics[height=6cm,width=6.5cm]{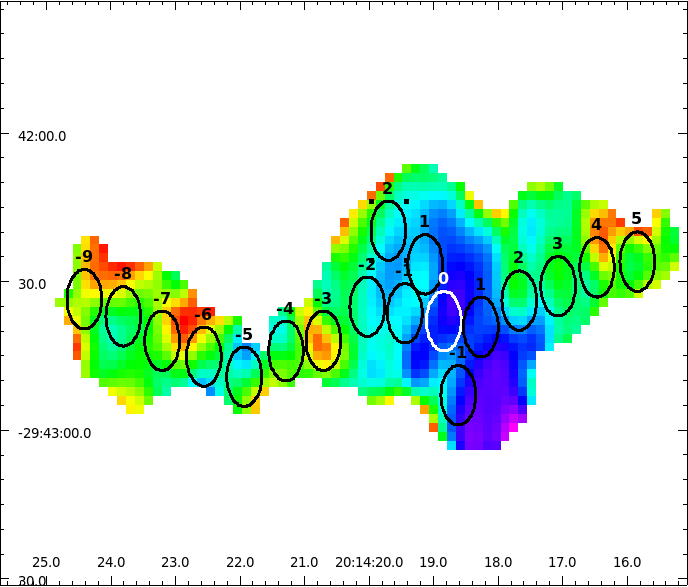}
	\smallskip \smallskip \smallskip
	\includegraphics[height=6cm,width=6.5cm]{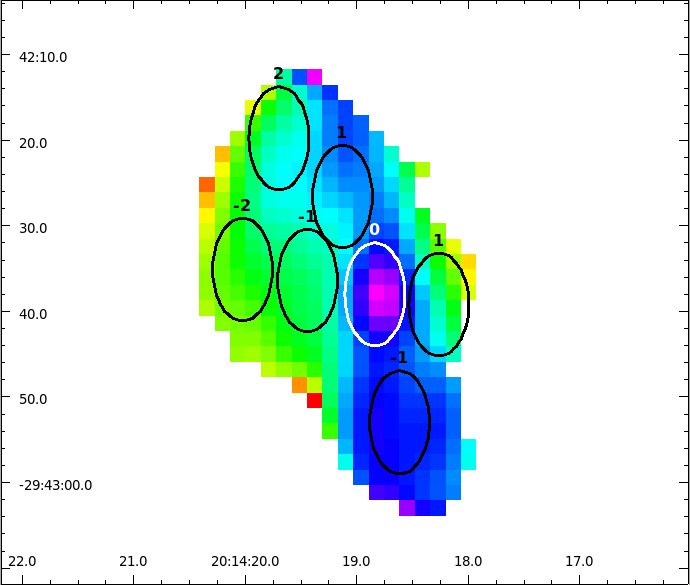}
	\smallskip \smallskip
	
	\includegraphics[height=5.5cm,width=17.5cm]{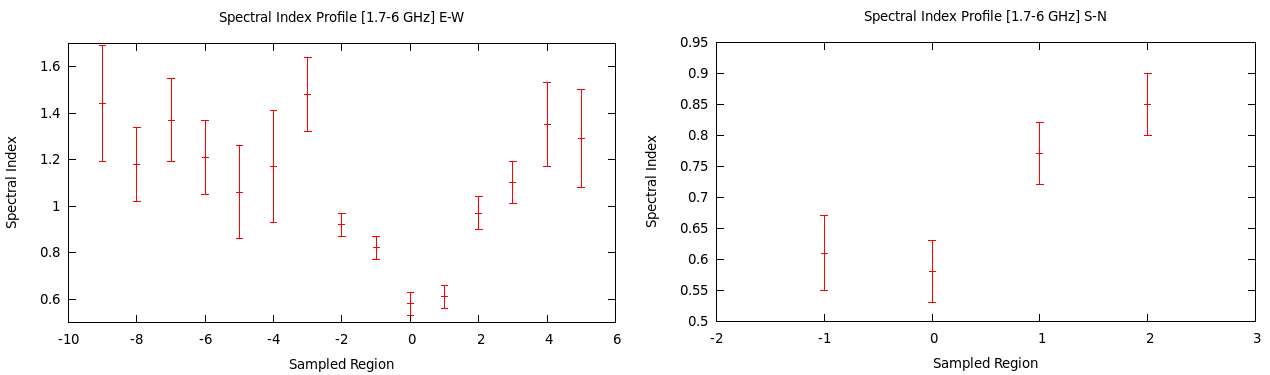}
	\smallskip \smallskip
	
	\includegraphics[height=5.5cm,width=17.5cm]{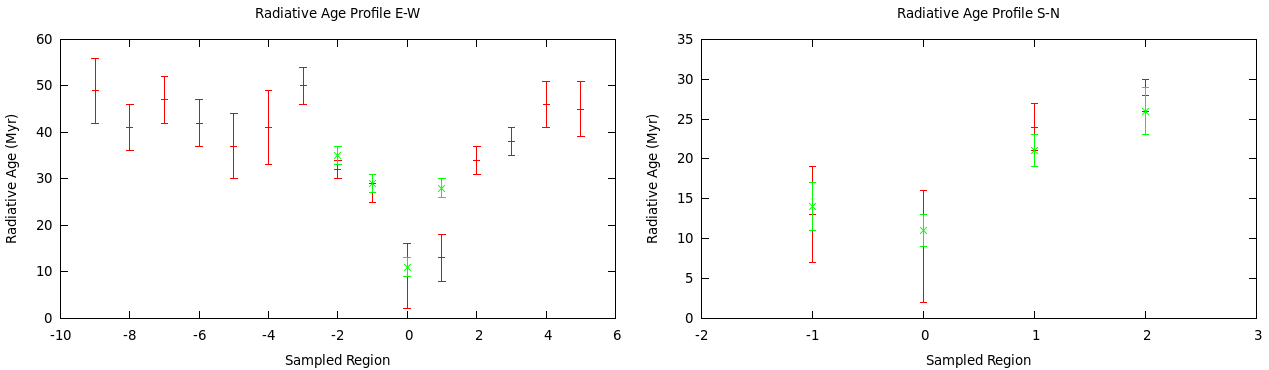} 
	\caption{{\it Top}: Sampling of the spectral index map between 1.7 and 6 GHz ({\it left}) and age map ({\it right}) with beam-size ($12''\times7''$) elliptical regions along the lobes and wings. The region '0' indicates the centre of the source. {\it Middle}: Spectral index profile along the wings ({\it left}) and primary lobes ({\it right}) derived from the sampling above. The index $\alpha$ steepens in both the directions, from the centre to the outer regions. {\it Bottom}: Radiative age profile along the wings ({\it left}) and the primary lobes ({\it right}). The red points represent the analytical age (Eq. \ref{formulaage}), the green points represent the values derived from the age map ({\ttfamily BRATS}). We note that they are consistent within $1\sigma$ errors. The age increases towards the outer parts of the source and we estimate a difference $\Delta t=22\pm7$ Myr between the ages of the wings and lobes.}
	\label{samplingspix}%
\end{figure*}

The radiative age maps were obtained with the Broadband Radio Astronomy ToolS software \citep[{\ttfamily BRATS}\footnote{http://www.askanastronomer.co.uk/brats/},][]{harwood,harwood2015}. {\ttfamily BRATS} computes age maps by combining radio images at different frequencies and compares them to ageing models through different statistical tests.  

We considered three ageing models: KP \citep{kardashev,pacholczyk}, JP \citep{jaffe} and Tribble \citep{tribble}. The KP model assumes a constant pitch angle (i.e. the angle between the electron velocity and the magnetic field vectors), the JP model considers instead a time-averaged pitch angle with respect to the radiative lifetime of an electron. Both models use a constant magnetic field. The Tribble model adopts JP-like radiative losses, but assumes also a Gaussian spatial distribution for the magnetic field. 
  
The input parameters necessary to the fit are the magnetic field $B$ (which is the average value of the Gaussian distribution in the case of the Tribble model) and the injection index $\Gamma$ (i.e. the initial spectral index). 
We used the 1.7, 5.5, and 9 GHz as input radio maps and we set $B=1.8$ $\mu$G (fixed as the equipartition magnetic field that we estimated for A3670) and $\Gamma=0.55$ (calculated by the {\ttfamily BRATS findinject} task).        

None of the models was rejected at the $68\%$ confidence level. The KP and JP models gave the highest $\chi_{\rm red}^2=1.44$ and the best $\chi_{\rm red}^2=1.21$, respectively, while the Tribble one gave an intermediate $\chi_{\rm red}^2=1.33$. The latter can describe a more general case than the KP and JP models, because the varying pitch angle and magnetic field appear more plausible than assuming constant values in every region of the source. Thus we consider the Tribble model as our best fit and the correspondent radiative age map obtained is shown in Fig. \ref{agemap}.  

The radiative age of the jets is $10\le t_{\rm rad}\le20$ Myr, while primary lobes are $20\le t_{\rm rad}\le30$ Myr old. The inner parts of the wings are $30\le t_{\rm rad}\le40$ Myr old. These results confirm the radial ageing along the wings direction and prove that the lobes are younger than the wings.   

The radiative age map in Fig. \ref{agemap} does not allow us to evaluate the outer parts of the wings at distances $>50$ kpc from the centre. However, as a first approximation, it is possible to use the following analytical expression (derived in Appendix A under simple assumptions) for the radiative age of a source observed at two frequencies as a function of the injection and spectral indices $\Gamma$ and $\alpha$:
\begin{equation}
t_{\rm rad}=A\frac{\sqrt{B}}{\left( B^2+B^2_{\rm CMB}\right)\sqrt{1+z}}\sqrt{\frac{\left(\alpha-\Gamma\right)\ln\left(\frac{\nu_1}{\nu_2}\right)}{\nu_1-\nu_2}} \; {\rm Myr}
\label{formulaage}
\end{equation}
where $B$ and $B_{\rm CMB}$$=3.25(1+z)^2$ (the equivalent Cosmic Microwave Background magnetic field) are expressed in $\mu$G, while the observing frequencies $\nu_1$ and $\nu_2$ are expressed in GHz. We adopt the constant $A=1590$ \citep[e.g.][]{slee,murgia}. 

In order to obtain a local spectral index in the outer regions, we sampled the spectral index map between 1.5 and 6 GHz with elliptical regions of the same beam sizes ($12''\times7''$), in both north-south and east-west directions, following the shape of the jets and the wings (Fig. \ref{samplingspix}, top left). The corresponding spectral index profiles (Fig. \ref{samplingspix}, middle) show that $\alpha$ steepens from region '0' (the centre of the source) to the outer regions of the wings and the lobes.  

In Eq. \ref{formulaage} we assumed $\Gamma=0.55$ and the local $\alpha$ from the sampling above (Fig. \ref{samplingspix}). As a first check, we verified that the estimated local age is consistent with that of the map in Fig. \ref{agemap}, by sampling it with the same regions (Fig. \ref{samplingspix}, top right). Then we produced the radiative age profiles shown in Fig. \ref{samplingspix}, bottom. The red points represent the analytical age from Eq. \ref{formulaage}, the green points represent the sampled age of the map (from {\ttfamily BRATS}). We note that they are consistent within $1\sigma$ errors. We observe an increasing ageing towards the outer parts of the source. In particular, the estimated ages of regions '-9' and '-1' are $t_{-9}=49\pm7$ Myr and $t_{-1}=27\pm2$ Myr. Therefore we conclude that the wings are $\Delta t=22\pm7$ Myr older than the lobes.

 We caution that equipartition conditions are ruled out by various inverse Compton observations, with the equipartition magnetic field $B_{\rm eq}$ likely being an overestimate of the true field strength $B_{\rm true}$ \citep[e.g.][]{ineson}. Owing to this, the spectral ages obtained above should be considered as lower limits for the true radiative ages. However, in the case of A3670, it is valid the inequality $B_{\rm true} < B_{\rm eq} < B_{\rm CMB}$, thus the inverse Compton losses dominate anyway and the true ages are not expected to be significantly different from the ones we estimate.

\subsection{The radio galaxy and its host properties}
\label{radio-ottico}
\begin{figure}
	\centering
	\includegraphics[height=6.5cm,width=7cm]{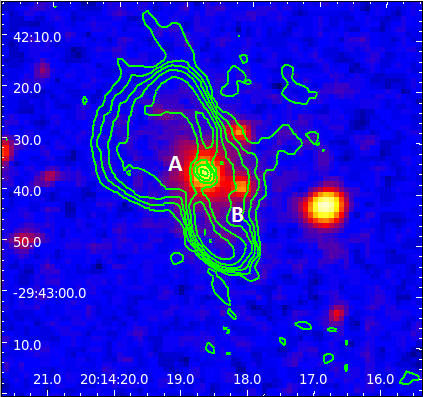}
	\smallskip
	
	\includegraphics[height=6.5cm,width=7cm]{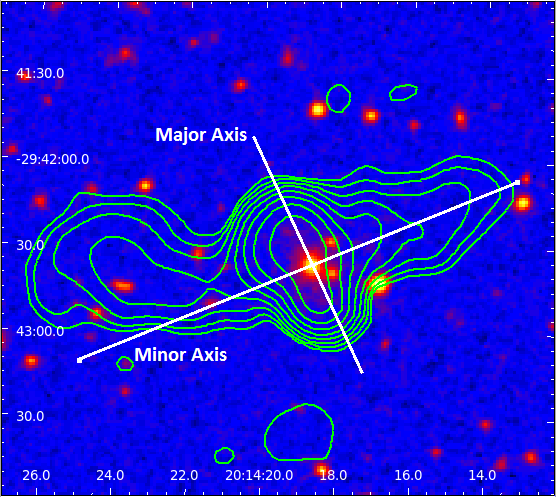} 
	\caption{{\it Top}: A3670 optical image (R-band, DSS) with overlaid 9 GHz radio contours. The primary and secondary optical nuclei are indicated with 'A' and 'B', respectively. No radio emission associated with the secondary nucleus is detected. {\it Bottom}: A3670 optical image (R-band, DSS) with overlaid 1.7 GHz radio contours. The white lines indicates the direction of the optical axes. The major axis is aligned with the primary lobes, while the minor one is nearly aligned with the wings.}
	\label{ottico}%
\end{figure}   

In order to compare the radio properties of A3670 to the optical properties of its host, in Fig. \ref{ottico} we overlay the radio contours at 1.7 GHz and 9 GHz to the R-band Digitized Sky Survey (DSS\footnote{http://archive.eso.org/dss/dss}) image of the host galaxy. 
As already mentioned, A3670 is a dumbbell galaxy. In the top panel of Fig. \ref{ottico}, the secondary optically fainter nucleus 'B' does not present radio emission and only the primary one 'A' is currently active. This result is in agreement with \citet{mchardy}, who found that only the brightest nucleus of a multiple optical system is usually active in the radio band.

The ellipticity and position angle of the host galaxy are $\varepsilon=0.28$ and $PA=24^o$ (measured from north to east). The cross in the bottom panel of Fig. \ref{ottico} indicates the direction of its major and minor optical axes. The major axis is aligned with the radio lobes, while the minor axis is nearly aligned with the wings. These results are consistent with \citet{capetti}, who finds the high ellipticity of the XRGs host galaxy and with \citet{gillone}, who reports radio-optical alignments similar to the ones we observed in A3670.    

The SMBH mass of the XRGs is found to be statistically higher than that of classical radio galaxies \citep{mezcua}. Several previous works \citep{kormendy,magorrian,ferraresemerritt} demonstrated that a strong correlation between the host and its central SMBH exists. We adopted these correlations to estimate the SMBH mass of A3670. By using the apparent magnitudes, we calculated the absolute magnitudes $M_{\rm B}=-21.75$ and $M_{\rm K}=-26.33$. Then we used the following mass-magnitude relations \citep{graham}:
\begin{equation}
\log\left({\frac{M_{\rm BH}}{M_\odot}}\right)=-0.40^{+0.05}_{-0.05} \left(M_{\rm B}+19.50\right)+8.27^{+0.08}_{-0.08}=9.17^{+0.03}_{-0.03}
\label{massaB}
\end{equation}  
\begin{equation}
\log\left({\frac{M_{\rm BH}}{M_\odot}}\right)=-0.33^{+0.09}_{-0.09} \left(M_{\rm K}+24\right)+8.33^{+0.15}_{-0.15}=9.10^{+0.06}_{-0.06}
\label{massaK}
\end{equation} 
Eq. \ref{massaB} and \ref{massaK} both give a SMBH mass of $M_{\rm BH}\sim10^9 \; M_{\odot}$, which is consistent with the results from the statistical analysis of \citet{mezcua}.

\section{Discussion}

In this section we briefly present the theoretical models proposed to explain the XRGs origin and we discuss their properties applied to the specific case of A3670 in order to determine the most plausible one. 

\subsection{Slow precession model}
In some cases, radio sources associated with a dumbbell galaxy exhibit primary and secondary lobes as typical XRGs, but their wings are off-set with respect to the centre and therefore these objects are classified as Z-shaped radio galaxies \citep{gopal2003}. This morphology could be induced by a slow precession of the jets, due to the tidal interaction with a companion galaxy \citep{wirth,dennettthorpe}. 

A precession caused by the interaction with the secondary optical nucleus may be responsible for the S-shape of the jets in A3670. However, we do not observe any off-set in the wings, therefore it is unlikely that this mechanism could have produced the wings, given the different morphologies of the X-shaped and Z-shaped sources.

\subsection{Reorientation models}
 In reorientation models, the spin of an active SMBH may abruptly change its direction ('spin-flip'). As a consequence, the jets are reoriented and the primary lobes evolve along the new direction, while the wings are fossil emission from previous jets. This phenomenon can occur after the coalescence with another SMBH \citep{merrittekers} or after the interaction between a SMBH (or a binary black hole system) and unstable regions of the accretion disk \citep{rees,dennettthorpe,liu}.

These models can explain why the wings never host jets and can also reproduce XRG morphologies with very extended wings. Moreover, they are consistent with a steep spectral index in the wings and with the high SMBH mass, which is caused by the coalescence or the presence of a non-resolved binary system. However, these models cannot explain the high ellipticity of the host galaxy and the observed radio-optical alignments \citep[e.g.][]{hodgeskluck2010,gopal}.

A3670 exhibits extended wings and a high SMBH mass that would be consistent with reorientation models. Nevertheless, they cannot account for the radio-optical alignments we find. The majority of the XRGs has an FRII morphology and FRI-type XRGs are rarer \citep{saripalli}. It is suggested that this evidence may be due to a different interaction between the SMBH and the accretion disk in the two classes of radio galaxies \citep{liu}. In fact, FRIIs typically have a radiatively efficient accretion disk, while FRIs have a radiatively inefficient one \citep{ghisellini}, which weakly interacts with the SMBH. This result makes the spin reorientation less likely and may well explain the lower number of FRI-type XRGs \citep{liu}. In the case of the XRG in A3670, which is an FRI, it appears implausible that a disk-SMBH interaction could have reoriented the jets by $\simeq90^o$, but we cannot exclude that a past coalescence may be the responsible.

\subsection{Hydrodynamical models}	

  Hydrodynamical models consider FRII-type jets aligned with the major axis of an high ellipticity galaxy, so that the environment gas pressure is stronger along the major axis with respect to the minor axis. The backflow plasma coming from the hotspots is therefore redirected towards the minor axis, where the minimum resistance of the gas allows the formation of the wings. In the buoyant backflow model \citep{leahywilliams,kraft} the wings plasma is led by the buoyancy force and evolves at subsonic speed. In a variant of this model \citep{capetti}, strong backflows may form a cocoon, which becomes over-pressured with respect to the surrounding gas and ejects plasma outflows at supersonic speed along the steepest pressure gradient (i.e. the minor axis), thus producing more extended wings. Three-dimensions numerical simulations suggest a supersonic origin and a subsonic evolution of the wings \citep{hodgeskluck2011}.

Hydrodynamical scenarios are supported by the direct observation of the high ellipticity of the host and the multiwavelength alignments in some well-studied XRGs \citep[e.g.][]{kraft}. The ellipticity and the radio-optical alignments in A3670 are consistent with these evidences.  A steep spectral index in the wings would be explained, because wings travel for a longer distance than the lobes. However, hydrodynamical models  cannot reproduce very extended wings, as those of A3670, and do not account for the high mass of its SMBH. Moreover, they usually require a FRII jet with hotspot in order to produce strong backflows driving the formation of the wings at subsonic or supersonic speed \citep{capetti,saripalli,hodgeskluck2011}. \citet{saripalli} consider the possibility of a jet evolution from FRII to FRI, that would make the wings to be produced in a past phase of activity when the jets still had hotspots, by a similar hydrodynamical process. The intermediate XRGs radio power would be a confirmation of this evolution. Thus, we cannot exclude these scenarios, but they can hardly explain the extended wings in A3670.

\subsection{Double AGN model}

 In the double AGN model, the lobes and the wings are independently produced by two active SMBH in a non-resolved binary system \citep{lalrao2007,lalrao2019}. 

This scenario can explain the XRGs with both steep and flat spectral indices in the wings. The estimated high mass of the XRG black hole is consistent with this model, because it would be the sum of the masses of two non-resolved SMBHs. Nevertheless, given the observed low occurrence of dual AGN \citep{burke}, it seems unlikely that the two SMBHs are active at the same time. Furthermore, this model cannot explain why the jets are never oriented along the wings and does not account for the alignments with the host galaxy. 

On the other hand, we cannot exclude that the primary optical nucleus of A3670 hosts a non-resolved SMBH binary system. It is possible that one of the two SMBHs had generated the wings in a previous phase of activity and then switched-off. After a time of $t\simeq20$ Myr (that is the estimated difference between the ages of the wings and the lobes), the other one switched-on and produced the lobes.

\subsection{Jet-stellar shell interaction model}

\begin{figure}
	\centering
	\includegraphics[height=6.5cm,width=7cm]{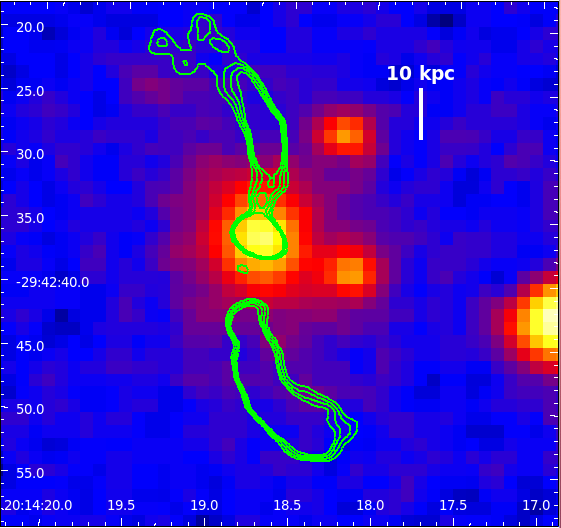}

	\caption{A3670 9 GHz map obtained with the {\ttfamily UNIFORM} weighting parameter (resolution is $2.0''\times1.2''$, RMS noise is 0.012 mJy$\cdot$ beam$^{-1}$; contour levels are $30\sigma$, $35\sigma$, $40\sigma$, $45\sigma$). The northern jet is at first deflected at $\simeq10$ kpc from the centre of the galaxy. }
	\label{uniform}%
\end{figure}   

According to the jet-stellar shell interaction model, the gas present in a stellar shell deflects the radio jets, thus producing the wings \citep{gopal}. This scenario predicts that a recent merger with a gas-rich disk galaxy has both activated the SMBH and produced a system of stellar shells. Stellar shells are rotating arc-shaped structures that have been found in $\sim10\%$ of local elliptical galaxies, roughly aligned with their optical major axis \citep{malin}. In particular, the interaction between the jets in \object{Centaurus A} and its shells at kpc scales \citep{gopalsaripalli} suggests that a similar phenomenon can occur in XRGs. For instance, in Centaurus A shells, both neutral and molecular hydrogen are found, with an estimated mass of $M_{\rm H}\simeq4\cdot10^7 \; M_\odot$ and an average density of $n_{\rm H}\simeq4\cdot10^{-2}$ cm$^{-3}$ \citep{gopal2003}.

Since the shells are aligned with the host major axis, the interaction can take place exclusively along this direction and the jets plasma is deflected towards the minor axis. In this model, the size of the wings depends on the duration of the interaction between the jet and the shells, which is determined by physical parameters like density and velocity, therefore very extended structures can be produced. Moreover, this scenario can account for high SMBH masses, because it is plausible to assume that a binary black hole system has formed or a coalescence has occurred, as a consequence of the galactic merging. Even if this model can reproduce the observed XRGs features, up to now stellar shells are confirmed only in two XRGs: \object{3C 403} \citep{almeida} and \object{4C+00.58} \citep{hodgeskluck2010b}.

Observations and numerical simulations suggest that BCGs are the result of several galaxy mergers \citep[e.g.][]{delucia,rasmussen}. The observed high SMBH masses and extended stellar halos in BCGs strongly support this formation scenario. A3670 exhibits the common properties of the dominant cluster galaxies and therefore a system of shells may have been produced after one of these mergers, as found in some other BCGs \citep{kluge}. In Fig. \ref{uniform} we report the radio contours of the high resolution ($2.0''\times1.2''$) 9 GHz map obtained with the {\ttfamily UNIFORM} weighting. We notice that the northern jet begins to be deflected at $\simeq 10$ kpc from the galaxy centre, on similar scales at which the shells are typically detected. Moreover, the rotational motion of the shells could explain the S-shape of the jets in A3670.

\section{Summary and conclusions}
In this work we investigated the properties and origin of the candidate XRG in the A3670 cluster. To this aim, we processed and analysed new JVLA radio data at 1.5, 5.5, 6, and 9 GHz. Here we summarise our results.

We obtained radio maps at different frequencies in order to accurately study the target. By considering the absence of hotspots in the primary lobes, we classify A3670 as an FRI-type XRG. We measure a 1.4 GHz radio power of $P_{1.4}=(1.7\pm0.1)\cdot10^{25} \; {\rm W\cdot Hz^{-1} }$. 
	
 The spectral index maps show a steepening from the centre to the outer regions of the source. The radiative age map suggests a progressive ageing from the centre, but it does not allow us to cover the whole source. Therefore, after checking its consistency with the age map, we used an approximate analytical expression to derive the radiative age as a function of the spectral index, which was measured out to the external regions. We estimate that the wings are $\Delta t=22\pm7$ Myr older than the lobes.
 
 Comparison between radio and optical data allowed us to verify some properties typically observed in XRGs. From literature data, we find a high ellipticity ($\varepsilon=0.28$) of the radio-loud component of the dumbbell system ('A' in Fig. \ref{ottico}). We report the alignments between the major and minor axes of the 'A' component with the radio lobes and wings, respectively. We estimate a high SMBH mass of $M_{\rm BH}\sim10^9\;M_{\odot}$.
 
Finally, we discussed the theoretical models proposed to explain the origin of XRGs. The considered models (slow precession, reorientation, hydrodynamical, and double AGN) cannot completely explain the observed properties of A3670, apart from the jet-stellar shell interaction one. Future multiwavelength data would allow us to better constrain the formation and evolution mechanism of the XRG in A3670, in particular further optical data are needed to confirm the presence of shells in the BCG of A3670.

\begin{acknowledgements}
We thank the referee for his comments and suggestions, that have significantly improved the presentation of the paper. LB thanks A. Ignesti for his useful suggestions. The National Radio Astronomy Observatory is a facility of the National Science Foundation operated under cooperative agreement by Associated Universities, Inc. This research has made use of the NASA/IPAC Extragalactic Database (NED), which is operated by the Jet Propulsion Laboratory, California Institute of Technology, under contract with the National Aeronautics and Space Administration. We acknowledge the usage of the HyperLeda database (http://leda.univ-lyon1.fr). Based on photographic data obtained using The UK Schmidt Telescope. The UK Schmidt Telescope was operated by the Royal Observatory Edinburgh, with funding from the UK Science and Engineering Research Council, until 1988 June, and thereafter by the Anglo-Australian Observatory. Original plate material is copyright (c) of the Royal Observatory Edinburgh and the Anglo-Australian Observatory. The plates were processed into the present compressed digital form with their permission. The Digitized Sky Survey was produced at the Space Telescope Science Institute under US Government grant NAG W-2166.  
\end{acknowledgements}

\bibliographystyle{aa}
\bibliography{bibliografia}

\appendix
\section{Radiative age-spectral index relation}
As it is generally known, the electronic density of a population of relativistic electrons follows a power law distribution in energy $n_{\rm e}(E)\propto E^{-\delta}$, where $\delta=2\Gamma+1$, and $\Gamma$ is the spectral index at the time $t=0$. Due to synchrotron and inverse Compton losses, the population becomes older and the spectrum steepens above a certain frequency ('break frequency', $\nu_{\rm b}$). The radiative age of the population can be estimated as \citep[e.g.][]{slee}:
\begin{equation}
t_{\rm rad}=A\frac{\sqrt{B}}{\left( B^2+B^2_{\rm CMB}\right)\sqrt{(1+z)\nu_{\rm b}}} 
\label{tradapp2}
\end{equation}  
The break frequency is generally not known a priori, but it is possible to easily find an expression of $t_{\rm rad}$ which is not explicitly dependent from it. Let $k=\frac{\nu}{\nu_{\rm b}}$ be defined as the ratio between the observing and the break frequencies. In the cases $\nu \simeq \nu_{\rm b}$ or $\nu > \nu_{\rm b}$, the synchrotron emissivity can be expressed as \citep{eilek}:  
\begin{equation}
J(k)=j_0n_{\rm e}BR_{\rm E}^{\delta-1}e^{-k}k^{-\frac{\delta-1}{2}}
\label{japp}
\end{equation}
where $j_0$ is the normalisation, and $R_{\rm E}$ is the ratio of the minimum and maximum energy of the electrons. 

Eq. \ref{japp} can be used to compute the spectral index between two frequencies $k_1$ and $k_2$:
\begin{equation*}
	\alpha=-\frac{\ln{\left( \frac{J_1}{J_2}\right)}}{\ln{\left( \frac{k_1}{k_2}\right)}}=- 
	\frac{\ln{\left(\frac{j_0n_{\rm e}BR_{\rm E}^{\delta-1}e^{-k_1}k_1^{-\frac{\delta-1}{2}}}{j_0n_{\rm e}BR_{\rm E}^{\delta-1}e^{-k_2}k_2^{-\frac{\delta-1}{2}}}\right)}}{\ln{\left( \frac{k_1}{k_2}\right)}}=-
	\frac{\ln{\left(\frac{e^{-k_1}k_1^{-\frac{\delta-1}{2}}}{e^{-k_2}k_2^{-\frac{\delta-1}{2}}}\right)}}{\ln{\left( \frac{k_1}{k_2}\right)}}=
\end{equation*}

\begin{equation}
=\frac{\left(k_1-k_2\right)}{\ln\left(\frac{k_1}{k_2}\right)}+\frac{\delta-1}{2}=
\frac{\left(k_1-k_2\right)}{\ln\left(\frac{k_1}{k_2}\right)}+\frac{2\Gamma+1-1}{2}=\frac{\left(\frac{\nu_1-\nu_2}{\nu_{\rm b}}\right)}{\ln\left(\frac{\nu_1}{\nu_2}\right)}+\Gamma
\label{alfapp}
\end{equation}
From Eq. \ref{alfapp}, one obtains the break frequency as a function of $\alpha$ and $\Gamma$:
\begin{equation}
\nu_{\rm b}=\frac{\nu_1-\nu_2}{\left(\alpha-\Gamma\right)\ln\left({\frac{\nu_1}{\nu_2}}\right)}
\label{break}
\end{equation}
Therefore, Eq. \ref{tradapp2} becomes:
\begin{equation}
t_{\rm rad}=A\frac{\sqrt{B}}{\left( B^2+B^2_{\rm CMB}\right)\sqrt{1+z}}\sqrt{\frac{\left(\alpha-\Gamma\right)\ln\left(\frac{\nu_1}{\nu_2}\right)}{\nu_1-\nu_2}}
\label{last}
\end{equation}


\end{document}